# Sulfur Plasma Opens Low-Temperature Reactive Sputtering to Diverse Sulfides

Daiki Motai[1], Issei Suzuki[1,*], Taichi Nogami[1], Takahisa Omata[1]

1. Institute of Multidisciplinary Research for Advanced Materials, Tohoku University, Sendai 980-8577, Japan

*Corresponding author: issei.suzuki@tohoku.ac.jp

## Abstract

Reactive sputtering underpins oxide and nitride thin-film technologies, but an equivalent approach for sulfides has remained elusive. Elemental sulfur vapor comprises mainly poorly reactive $S_8$ molecules, whereas $H_2S$ is highly toxic and can introduce hydrogen impurities. Here, sulfur plasma generated from elemental sulfur provides a $H_2S$-free platform for extending reactive sputtering to sulfides. The plasma supplies dissociated sulfur species carrying high chemical energy to the growth surface at a rate controlled independently of the metal supply. These characteristics enable low-temperature crystal growth without post-annealing and control over film composition, while maintaining high deposition rates of metal atoms, typically 60-80% of those without sulfur supply. Using a single sulfur plasma source with W, Mo, Zn, and Cu metallic targets yields layered $WS_2$ and $MoS_2$ films grown at 300 °C with their van der Waals planes parallel to the substrate, including $MoS_2$ on polyimide film. ZnS films exhibit reduced sulfur deficiency through adjustment of the metal-to-sulfur supply ratio, while $Cu_{1.8}S$ is deposited on unheated substrates. Sulfur plasma thus provides a reactive sulfur source analogous to oxygen and nitrogen radical sources in oxide and nitride deposition, extending the versatility of reactive sputtering to diverse sulfide thin films.

## 1. Introduction

Oxide and nitride thin films are widely fabricated by reactive sputtering, in which metallic targets are sputtered in reactive $O_2$ and $N_2$ atmospheres. This technique underpins the production of transparent conductive oxide films, hard nitride coatings, and piezoelectric films [1,2]. By using readily available metallic targets rather than compound targets, reactive sputtering also enables control over film composition through adjustment of the metal and reactive gas supplies. Furthermore, supplying active species to the growth surface from oxygen or nitrogen radical or plasma sources enables the growth of highly crystalline films even at low temperature [3]. Sulfides, in contrast, lack an equivalent approach. Elemental sulfur vapor consists mainly of stable $S_8$ molecules and does not function as a reactive gas in the way that $O_2$ and $N_2$ do in conventional reactive sputtering: $S_8$ molecules react inefficiently at the growth surface [4]. Sulfurization using sulfur vapor generally relies on a large excess of sulfur and high substrate temperatures [5]. Reactive sputtering using $H_2S$ gas has also been reported [6–8]. However, $H_2S$ is highly toxic and requires strict safety management. Moreover, hydrogen supplied with sulfur can be incorporated into the film as an impurity, complicating the control of film properties [9,10]. Sulfurization of the target surface (target poisoning) also reduces the deposition rate [6,11]. Consequently, sulfide thin films have primarily been fabricated by sputtering from sintered sulfide targets, chemical vapor deposition (CVD), or sulfurization of metal or metal oxide precursor films. These approaches often require high processing temperatures to promote crystallization or sulfurization [12,13].

Sulfides encompass a wide range of functional materials. Examples include ZnS, which is used in phosphors and optical applications; CuxS, SnS, and CuInS2, which have been investigated as absorber materials for solar cells; and transition metal dichalcogenides (TMDs), such as WS2 and MoS2, which have attracted considerable attention in recent years as two-dimensional electronic materials [14–16]. To scale these materials up to large areas, integrate them into multilayer structures with dissimilar materials, or fabricate flexible devices on polymer substrates, the entire process, up to obtaining films with the intended crystal structure and composition, must be completed at low temperature to limit thermal effects on the substrate and underlying films [17]. Thus, reactive sputtering of sulfides faces two key challenges: growing crystals of the intended phase at low temperature and controlling the sulfur content of the deposited films.

The first challenge is achieving crystal growth at low temperatures. In layered TMDs, for example, exploiting the properties arising from their two-dimensional nature requires films with the intended crystalline phase and sufficiently developed well-oriented layers. Such films have typically required high processing temperatures. For example, $MoS_2$ films with layers nearly parallel to the substrate have been obtained by sulfurizing $MoO_3$ precursor films at 800 °C [13]. $MoS_2$ films sputtered from a compound target crystallized at a substrate temperature of 400 °C but remained nanocrystalline and exhibited poor preferred orientation [18]. Achieving well-oriented layered crystals without high substrate temperatures or post-deposition treatment therefore requires an alternative source of energy for crystal growth and orientation. One promising approach is to deliver active sulfur species carrying high chemical energy directly to the growth surface, analogous to the use of radical sources in oxide and nitride deposition.

The second challenge is controlling the sulfur content of the deposited film. During vacuum deposition, supplied sulfur tends to re-evaporate before being incorporated into the film, resulting in sulfur deficiency. In $MoS_2$, sulfur deficiency results in an unintended increase in the n-type carrier density [19], while in ZnS, it degrades optical transparency [20]. In systems that contain multiple phases with different compositions, such as Sn-S and Cu-S, sulfur deficiency can also promote the formation of undesired phases [21,22]. Obtaining films with the intended composition and physical properties therefore requires independent control of the sulfur supply relative to the metal supply.

Taken together, these challenges call for a sulfur source that can supply chemically active sulfur species without using $H_2S$ and whose sulfur supply can be adjusted independently of the metal source. Radio-frequency (RF) plasma activation of elemental sulfur vapor dissociates $S_8$ molecules into species such as $S_2$ molecules, S atoms, and S ions [5,23]. The resulting plasma serves as a chemically active sulfur source [4]. Combining this sulfur plasma with metallic-target sputtering enables $H_2S$-free reactive sputtering, with metal and sulfur supplied simultaneously to the growth surface from independent sources. In our previous study, this approach yielded stoichiometric SnS films with large crystal grains at low temperature [24], demonstrating the feasibility of sulfur plasma as a reactive sulfur source for sulfide deposition.

In this work, we deposit thin films of four sulfides that differ in crystal structure and chemical nature from W, Mo, Zn, and Cu metallic targets using a single sulfur plasma source. Using these sulfides as case studies, we show three characteristics that sulfur plasma brings to film deposition: crystal growth at low temperature enabled by active sulfur species, control of the sulfur content of the films through a sulfur supply independent of the metal supply, and a small decrease in the deposition rate upon sulfur supply (i.e., suppressed target poisoning). These results establish that sulfur plasma opens reactive sputtering to diverse sulfides, playing for them the role that oxygen and nitrogen play for oxides and nitrides.

## 2. Results and discussion

Figure 1 shows the configuration of the deposition apparatus used in this work. A metallic target is mounted on a pressure-gradient sputtering (PGS) cathode, through which sputtering gas (Ar) is introduced in the vicinity of the target. This cathode and a sulfur plasma source are installed in the same chamber, enabling metal and sulfur to be supplied simultaneously to the substrate (Figure 1(a)). In the sulfur plasma source, sulfur vapor is introduced together with Ar into an inductively coupled plasma (ICP), where $S_8$ molecules are dissociated. The resulting sulfur species are then directed toward the substrate (Figure 1(b)) [25,26]. The distance from the sulfur plasma source outlet to the substrate is 85 mm, and that from the target to the substrate is 135 mm. The metal and sulfur supplies can be controlled independently: the former by the RF power applied to the target (RFMetal) and the latter by the heating temperature of the elemental sulfur ($T_{Sulfur}$).

Table 1 summarizes the sulfide thin films deposited in this work by reactive sputtering with sulfur plasma. The W, Mo, Zn, and Cu metallic targets yielded layered $WS_2$ and $MoS_2$, wurtzite-type ZnS, and digenite-type $Cu_{1.8}S$, respectively (for $Cu_{1.8}S$, see Section S4 in the Supporting Information). Together with SnS, reported previously [24], this brings the total number of investigated sulfides with differing crystal structure and chemical nature to five. All the sulfides were obtained as crystalline thin films from a single sulfur source at substrate temperatures from room temperature to 300 °C. In subsequent sections, we use these films to show the three key advantages of sulfur plasma: low-temperature crystal growth (Section 2.1), control of film composition (Section 2.2), and suppressed target poisoning (Section 2.3).

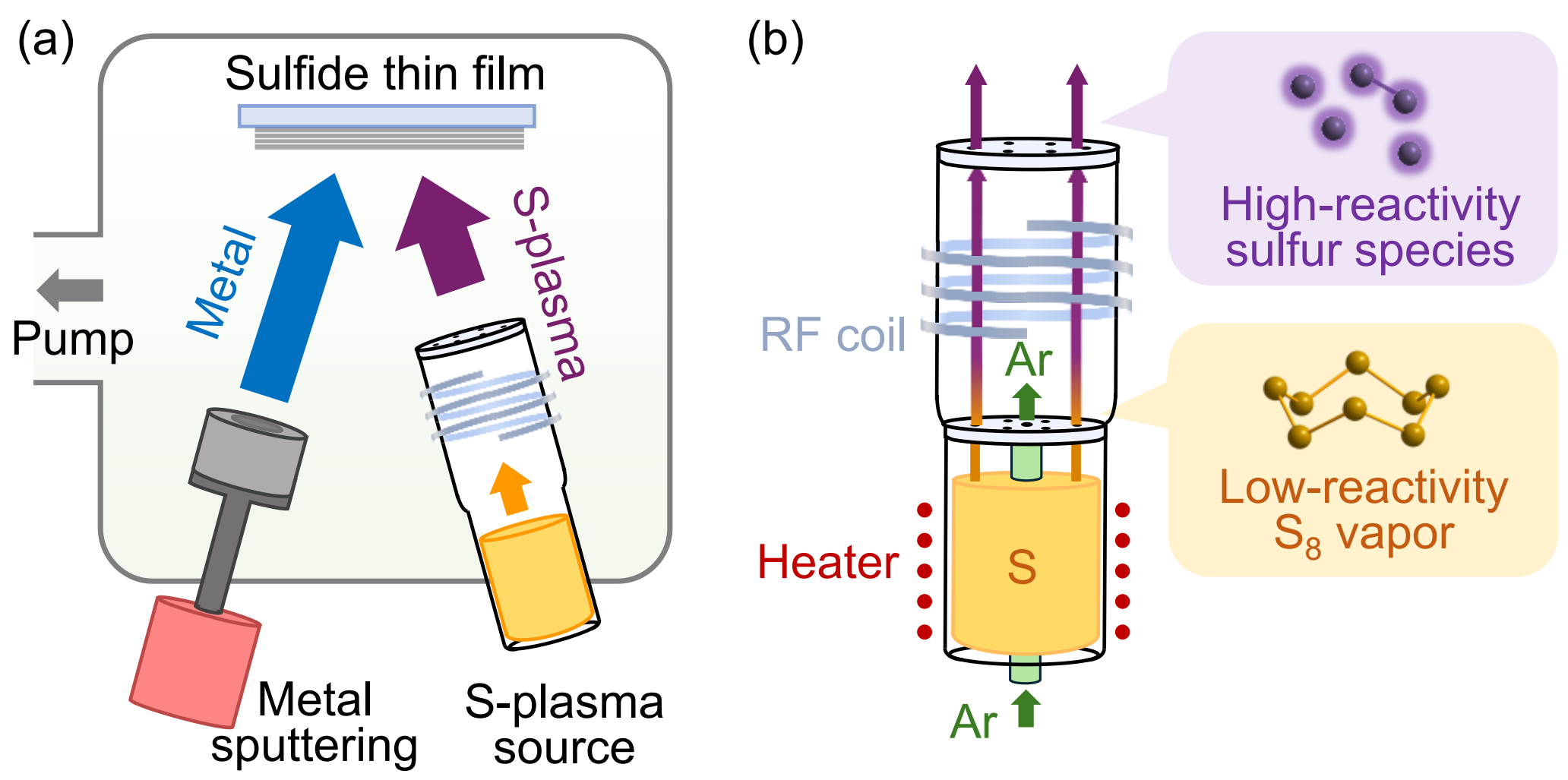

**Figure 1.** Schematic illustrations of reactive sputtering with sulfur plasma. (a) Deposition configuration in which sputtered metal species and reactive sulfur species are simultaneously supplied to the substrate for sulfide thin-film growth. (b) Sulfur plasma source, where low-reactivity $S_8$ molecules in sulfur vapor are excited and dissociated by an inductively coupled plasma to generate active sulfur species.

**Table 1.** Sulfide thin films deposited by reactive sputtering using sulfur plasma in this work, together with the deposition conditions (substrate temperature $T_{Sub}$, RF power applied to the metallic target $RF_{Metal}$, and heating temperature of the sulfur source $T_{Sulfur}$) and the crystal phases. The other conditions were common to all depositions and are given in the Experimental Section. SnS, deposited with the same apparatus in our previous work, is included for comparison.

| Metal target | Sulfide film | Crystal structure | $T_{Sub}$ / °C | $RF_{Metal}$ / W | $T_{Sulfur}$ / °C | Data |
|---|---|---|---|---|---|---|
| W | $WS_2$ | 2H-type (layered) | 300 | 5 | 105 | Figure 2 |
| Mo | $MoS_2$ | 2H-type (layered) | 300 | 5 | 100 | Figure 2 |
| Zn | ZnS | wurtzite-type | RT[a)] | 30/5 | 100/105 | Figure 3 |
| Cu | $Cu_{1.8}S$ | digenite-type | RT[a)] | 10 | 100 | Figure S3 |
| Sn | SnS | α-SnS type (layered) | RT-300 | 5 | 100 | Ref.[b)] |

a) RT: unheated substrate (room temperature)
b) Data from Ref.[24]

## 2.1 Low-temperature crystal growth: layered $WS_2$ and $MoS_2$

Thin films were deposited on $SiO_2$ glass substrates at 300 °C by sputtering W and Mo metallic targets under a sulfur plasma supply, without post-deposition annealing. Figure 2(a) shows a cross-sectional high-resolution transmission electron microscopy (HRTEM) image of the film deposited using the W metallic target. Layered contrast parallel to the substrate is clearly observed throughout the film, with an interlayer spacing of 7.0 Å. The fast Fourier transform (FFT) pattern obtained from this region of the film (Figure 2(b)) shows spots corresponding to the 002, 004, and 006 reflections of $WS_2$. The interlayer spacing determined from the spot positions is 7.1 Å, in good agreement with the value obtained from the real-space image.

Figure 2(c,d) shows the out-of-plane and in-plane X-ray diffraction (XRD) patterns of the films deposited using the W and Mo metallic targets, respectively. In the out-of-plane pattern of the film deposited from the W target, only the 002 reflection of hexagonal 2H-$WS_2$ is observed, as a strong peak at $2\theta = 12.6°$. The in-plane pattern shows the 100, 110, and 200 reflections, which arise from (hk0) planes perpendicular to the substrate. The interplanar spacing derived from the 002 reflection is 7.0 Å, consistent with the interlayer spacings obtained from the HRTEM image (7.0 Å) and the FFT pattern (7.1 Å). Similarly, the film deposited from the Mo target exhibits only the 002 reflection at $2\theta = 13.8°$ in the out-of-plane diffraction pattern and only the 100, 110, and 200 reflections in the in-plane diffraction pattern. These results indicate that both the $WS_2$ and $MoS_2$ films exhibit a $00\ell$ orientation,

with their van der Waals planes aligned parallel to the substrate. The amorphous $SiO_2$ glass substrates do not provide an epitaxial template; hence, the observed orientation in both films is attributed to self-orientation driven by the low surface energy of the van der Waals planes [27,28]. In both films, the 002 reflections occur at lower diffraction angles than the corresponding bulk reflections, whereas the in-plane reflections occur at higher angles. These shifts correspond to an expansion of the c-axis lattice parameter by approximately 13% for $WS_2$ and 4% for $MoS_2$, and a contraction of the a-axis lattice parameter by approximately 4% and 1%, respectively, determined from the in-plane reflections. Such c-axis expansion has been reported previously [6,29] and is attributed to turbostratic stacking, in which adjacent layers are stacked with in-plane rotational and translational misalignment [6,30] (see Section S1 in the Supporting Information).

As described in the Introduction, the challenge in low-temperature deposition of TMD films is not only crystallization but also sufficient growth of the layers and parallel alignment to the substrate during deposition. For sputter deposition of $WS_2$, Villamayor et al. used a $WS_2$ target in an $Ar/H_2S$ atmosphere and reported that crystallinity and orientation improved with increasing substrate temperature; obtaining films with layers parallel to the substrate required a temperature as high as 700 °C [8]. Crystalline TMD films have been obtained at 300 °C by atomic layer deposition (ALD) using a Mo precursor and $H_2S$ [31] and by sulfurization of metallic Mo/W precursor films with a $H_2S/Ar$ plasma [32]. However, all these methods rely on $H_2S$ as the sulfur source, posing toxicity concerns and introducing hydrogen impurities. In contrast, the present method uses sulfur plasma generated from elemental sulfur vapor to produce $00\ell$-oriented $WS_2$ and $MoS_2$ films at 300 °C without post-deposition annealing, avoiding the use of $H_2S$.

The direct growth of layered crystals at low temperature in the present method is assisted by the energy delivered to the growth surface along with active sulfur species. In sputter deposition, not only the substrate temperature but also the kinetic and potential energies of the particles arriving at the growth surface affect surface atomic migration and crystal growth [33]. In the present method, the potential energy of the sulfur species is transferred to the growth surface in the following three ways. The first is the energy released during sulfide formation from sulfur species with high chemical energy. Based on the standard enthalpies of formation at 298 K, S atoms and $S_2$ molecules have enthalpies that are higher by 2.7 eV and 0.54 eV per sulfur atom, respectively, than those of the $S_8$ molecules that dominate sulfur vapor [34]. Per formula unit of $WS_2$, the formation reaction therefore releases approximately 1.1 eV more energy when sulfur is supplied as $S_2$ molecules and approximately 5.5 eV more when it is supplied as S atoms than when it is supplied as $S_8$. The second is the recombination, on the growth surface, of excess sulfur that is not incorporated into the film. Based on the sulfur supply estimated from the weight loss of the sulfur source [24], only 1–3% of the supplied sulfur is incorporated into the $WS_2$ and $MoS_2$ films. Hence, the region near the growth surface contains a large excess of sulfur species. For example, when S atoms that have arrived at the growth surface recombine and desorb as $S_2$ molecules, an energy corresponding to the $S_2$ bond energy (4.4 eV [34]) is released and shared between the growth surface and the desorbing $S_2$ molecules, which carry part of it away as kinetic energy [35]. The third is the neutralization of sulfur ions. The sulfur plasma also contains a fraction of ions such as $S^+$ [5,23]. When an $S^+$ ion is neutralized at the growth surface, an energy corresponding to the difference between the ionization energy of S (10.4 eV [36]) and the work function of the film is delivered to the surface [35]. With the work functions of $WS_2$ and $MoS_2$ (~4.5 eV and ~4.8 eV, respectively [37]), each neutralization of $S^+$ delivers 5.9 eV to $WS_2$ and 5.6 eV to $MoS_2$. These energies promote atomic diffusion and rearrangement on the growth surface. We attribute the growth of layered crystals at 300 °C, with their low-surface-energy van der Waals planes parallel to the substrate, to this enhanced surface diffusion.

Moreover, because active sulfur species are supplied to the growth surface simultaneously with the metal species, the present method does not require sulfur to diffuse through the entire deposited film, as post-deposition sulfurization does. This may bring an additional benefit, namely a sulfur concentration that is more easily kept uniform across the film thickness.

In addition, a $MoS_2$ film was deposited on a flexible polyimide substrate under the same conditions as on the $SiO_2$ glass substrates. No visible delamination occurred when the film was bent by hand (Figure 2(e)). The entire process is completed at low temperature; hence, the present method is applicable to polymer substrates and can be used to fabricate TMD films directly on flexible substrates without a transfer step. These results show that reactive sputtering with sulfur plasma promotes crystallization and orientation, resulting in the growth of oriented layered TMD films at 300 °C, a lower temperature than conventional methods require, without incorporating hydrogen impurities.

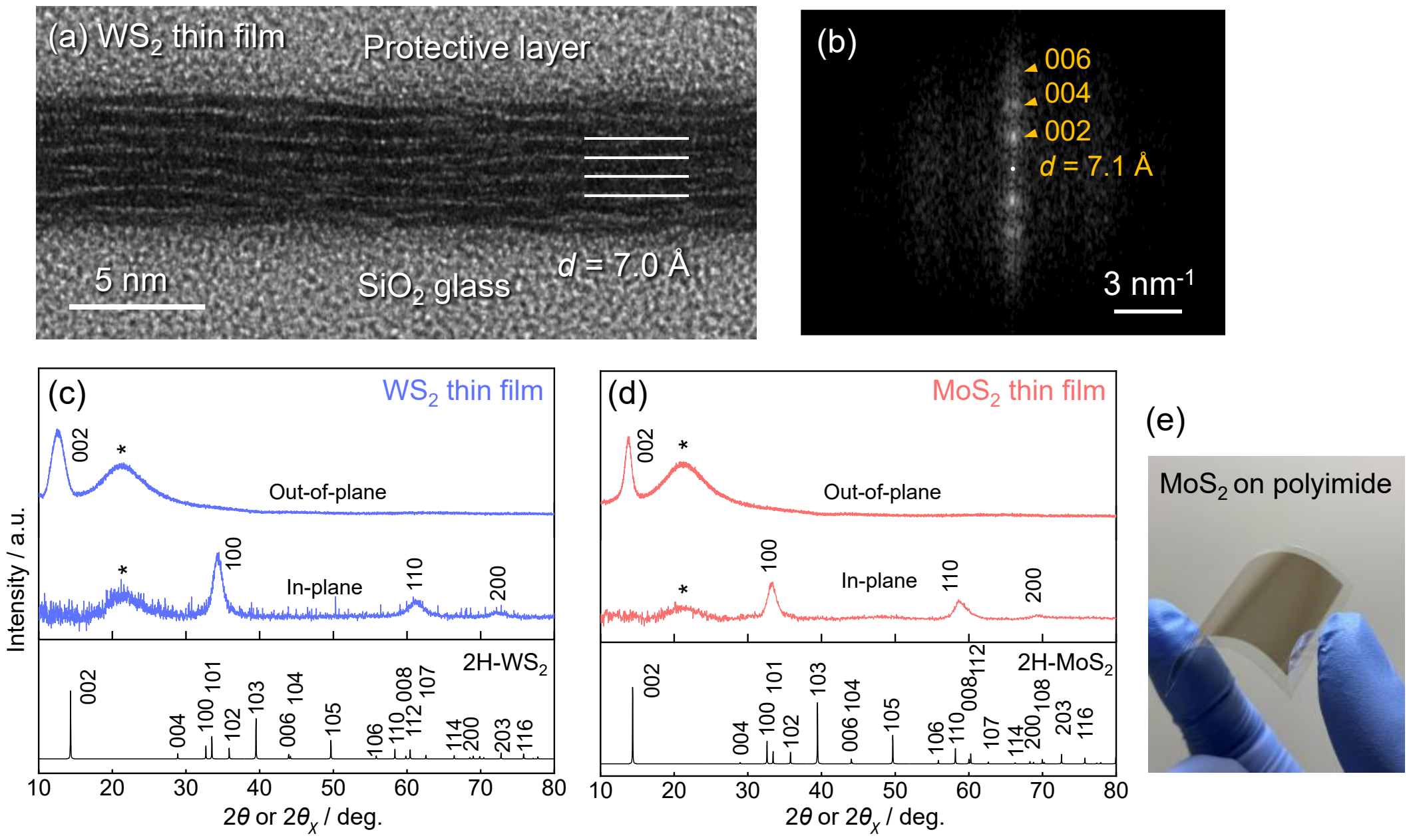

**Figure 2.** (a) Cross-sectional HRTEM image of the $WS_2$ thin film deposited at 300 °C. (b) FFT pattern obtained from the thin-film region of the HRTEM image in (a). (c,d) Out-of-plane and in-plane XRD patterns of the $WS_2$ and $MoS_2$ thin films, together with the powder patterns of 2H-$WS_2$ (ICSD#204265 [38]) and 2H-$MoS_2$ (ICSD#95569 [39]), respectively. The broad peaks marked with asterisks originate from the $SiO_2$ glass substrates. The deposition times were 30 min for the $WS_2$ film and 1 h for the $MoS_2$ film. (e) Photograph of a $MoS_2$ thin film deposited directly on a polyimide substrate under the same conditions as the film in (d), taken while the film was bent by hand.

### 2.2 Controllability of chemical composition: ZnS

The effects of the metal and sulfur supply conditions on the film composition and properties were examined by depositing ZnS films on unheated $SiO_2$ glass substrates. Figure 3(a) shows the out-of-plane and in-plane XRD patterns of the film deposited at RFZn (RFMetal for the Zn target) = 30 W and $T_{Sulfur}$ = 100 °C. The strong peak at 2θ = 29° in the out-of-plane pattern can be assigned to either the 111 reflection of zincblende-type ZnS or the 002 reflection of wurtzite-type ZnS. The in-plane pattern, however, shows the 100 and 101 reflections of the wurtzite type in an angular range where the zincblende type has no reflections, identifying the film as wurtzite-type ZnS. The thermodynamically stable phase of ZnS at room temperature is the zincblende type, whereas the wurtzite type is the high-temperature phase [40]. In contrast, a reference film sputtered from a ZnS target at 200 °C without sulfur plasma was a mixture of wurtzite and zincblende types (see Section S3 in the Supporting Information for details). The formation of the high-temperature phase with sulfur plasma supports the conclusion in Section 2.1 that sulfur plasma delivers energy to the growth surface.

The composition determined by electron probe microanalysis (EPMA) was S/Zn = 0.71, a large deviation from stoichiometry toward sulfur deficiency. As shown in Figure 3(c), this film had a markedly low transmittance in the visible range and was nearly black in appearance. To increase the sulfur supply relative to the Zn supply, RFZn was then lowered from 30 W to 5 W and $T_{Sulfur}$ was raised from 100 °C to 105 °C. The film obtained under these conditions was likewise identified as wurtzite-type ZnS (Figure 3(b); surface and cross-sectional scanning electron microscopy images in Figure S1 in the Supporting Information show that both films are flat and dense). Its composition was S/Zn = 0.91, closer to the stoichiometric composition. In addition, the transmittance in the visible range increased substantially, and the film became transparent in appearance (Figure 3(c)). This results from the reduced density of defects caused by sulfur deficiency [20,41]. However, the film with S/Zn = 0.91 is still sulfur-deficient. Because the metal and sulfur supplies in this apparatus can be further adjusted through the target-to-substrate distance and the sulfur heating temperature [24,42], stoichiometric ZnS films may also be achievable.

These results show that the present method enables the film composition to be tuned by adjusting the metal and sulfur supplies independently. This is a major advantage relative to existing methods, as it enables the deposition of various compounds that differ in their bonding energy with sulfur.

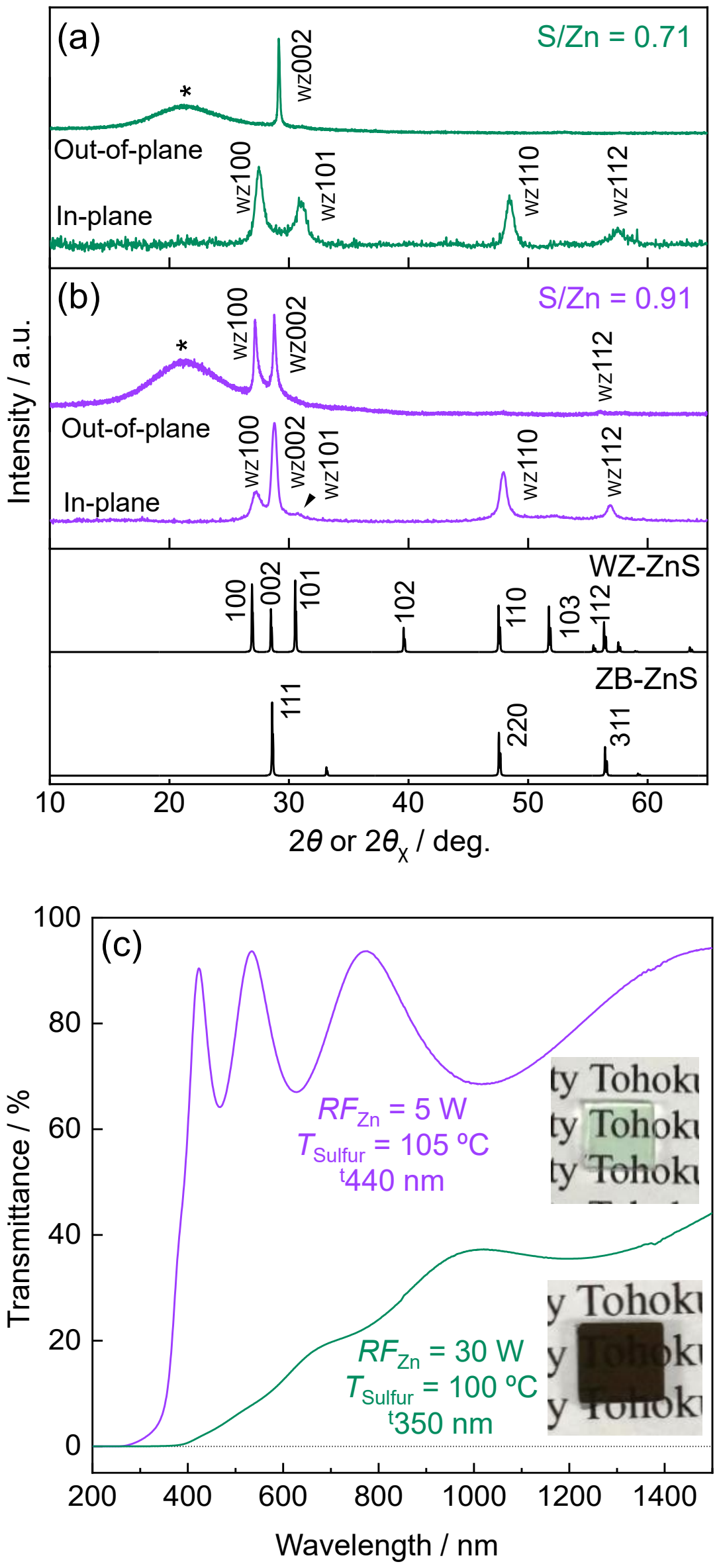

**Figure 3.** Out-of-plane and in-plane XRD patterns of the ZnS thin films deposited on unheated $SiO_2$ glass substrates at (a) RFZn = 30 W and $T_{Sulfur}$ = 100 °C (S/Zn = 0.71) and (b) RFZn = 5 W and $T_{Sulfur}$ = 105 °C (S/Zn = 0.91), together with the powder patterns of wurtzite-type (WZ: ICSD#230703 [43]) and zincblende-type ZnS (ZB: ICSD#67453 [44]). Reflections labeled "WZ" are indexed to the wurtzite-type structure. The broad peaks marked with asterisks originate from the $SiO_2$ glass substrates. (c) Optical transmission spectra of the films shown in (a) and (b). The insets show photographs of the thin films.

## 2.3 Suppressed target poisoning

In reactive sputtering, the target surface often reacts with the reactive gas and becomes covered with a compound (target poisoning), which lowers the deposition rate. The state in which the target surface remains close to metallic is called the metallic mode, and the state in which it is covered with the compound is called the compound mode. For oxides, the deposition rate in the compound mode drops to approximately one-tenth of that in the metallic mode [2,45,46]. To evaluate the extent of target poisoning under sulfur plasma supply, the metal-atom deposition rates were compared, for each metal, between metallic films deposited without sulfur supply and sulfide films deposited with sulfur plasma (Table 2; detailed deposition rates are listed in Table S2 in the Supporting Information). Except for Zn, the metal-atom deposition rates in the sulfide films were 0.6–0.8 times those in the metallic films. This means that 60–80% of the metal atoms that would be deposited as a metallic film are still incorporated into the film as a sulfide under sulfur plasma supply. For Zn, the rate was instead higher in the sulfide film by a factor of 1.7. This is likely because metallic Zn, with its high vapor pressure, condenses poorly as a metallic film at deposition rates as low as those used in this study, whereas under sulfur plasma supply it is readily incorporated into the film as ZnS, which has a considerably lower vapor pressure (see Section S6 in the Supporting Information). For W, Mo, and Sn, the ratios obtained with the present method are several times higher than those of about 0.1–0.3 estimated from reports on reactive sputtering of sulfides with $H_2S$ (Table 2) [6,11,47]. Target poisoning is thus suppressed even under sulfur plasma supply.

Target poisoning is suppressed because only a small amount of sulfur reaches the target surface. $H_2S$ fills the chamber as a gas and acts on the target at the same partial pressure as on the growth surface. In the present method, by contrast, the outlet of the sulfur plasma source is closer to the substrate than the metallic target is. In addition, the PGS cathode keeps the pressure near the target high [48] and creates a gas flow from the cathode through the deposition zone toward the exhaust port, so sulfur in the deposition zone is less likely to flow toward the target. Furthermore, unlike gaseous $H_2S$, elemental sulfur is solid at room temperature. Excess sulfur not incorporated into the film therefore largely condenses on chamber walls and other surfaces rather than accumulating in the chamber atmosphere. Although sulfur is supplied in large excess in the present method, with only about 1–3% incorporated into the film, the excess sulfur is lost through these processes before it reaches the target. As a result, the target remains close to the metallic mode. Because the sulfur partial pressure and target poisoning are not inextricably linked as they are with $H_2S$, the present method is inherently suited to achieving high deposition rates.

**Table 2.** Ratios of the deposition rate of metal atoms in sulfide films to that in metallic films deposited without sulfur supply, for reactive sputtering with sulfur plasma (this work) and with $H_2S$ (reported). For this work, metal-atom counts were estimated from film thicknesses assuming bulk densities of the respective phases. The reported values were obtained with different apparatus and conditions.

| Metal / sulfide | Metal in sulfide film / metal in metallic film | |
|---|---|---|
| | Sulfur plasma (this work) | $H_2S$ (reported) |
| Cu / $Cu_{1.8}S$ | ~0.8 | – |
| W / $WS_2$ | ~0.7 | ~0.1 a) |
| Mo / $MoS_2$ | ~0.6 | ~0.3 b) |
| Sn / SnS | ~0.7 c) | ~0.25 d) |
| Zn / ZnS | ~1.7 e) | – |

–: not reported.

a) Estimated from [6]; the value lies between those obtained from film thicknesses assuming bulk densities (~0.05) and from the W Lα fluorescence intensities (~0.15).

b) Estimated from [47] assuming bulk densities and the same thickness (≈20 nm for 30 min, as stated for the film series) for the Mo and $MoS_2$ films.

c) Data from [24].

d) Estimated from film thicknesses in [11] assuming bulk densities.

e) Exceeds 1 because the deposition rate of the metallic Zn reference film is lowered by the low condensation coefficient of metallic Zn (see Section S6).

## 3. Conclusion

This work has shown that reactive sputtering with sulfur plasma from a single sulfur source yields sulfide thin films that vary in crystal structure and chemical nature. This generality stems from three characteristics of sulfur plasma. First, the plasma delivers dissociated and partly ionized sulfur species, and the energy they release at the growth surface through reaction, recombination, and neutralization drives crystal growth at low temperatures, as demonstrated by $WS_2$ and $MoS_2$, which grow at 300 °C as layered crystals with their van der Waals planes parallel to the substrate. Second, because sulfur is supplied independently of the metal, the sulfur content of the film can be controlled, as demonstrated by ZnS, whose sulfur deficiency and hence optical transmission were tuned by adjusting the metal-to-sulfur supply ratio. Third, sulfur is delivered locally to the growth surface, and the excess condenses as a solid instead of accumulating as a background gas, so that target poisoning is suppressed: the metal-atom deposition rates of the sulfide films typically remain at 0.6–0.8 of those of the metallic films, several times the values reported for $H_2S$-based reactive sputtering.

For oxides and nitrides, reactive sputtering using oxygen and nitrogen gases or radicals is widely established, from laboratory research to production lines. Despite the similarly diverse functionalities of sulfides, a comparable deposition approach has yet to be established, largely because a reactive, safe, and controllable sulfur source is lacking. With sulfur plasma providing this source, reactive sputtering can serve as a universal deposition method for sulfide thin films as well.

## 4. Experimental Section

**Thin-Film Deposition:** ZnS, $WS_2$, $MoS_2$, and $Cu_{1.8}S$ thin films were deposited on $SiO_2$ glass substrates (40 mm x 40 mm, 1 mm thick) by reactive sputtering with sulfur plasma. $MoS_2$ thin films were also deposited on a 50-μm-thick transparent polyimide film (Neopulim S100, Mitsubishi Gas Chemical Company, Inc., Japan) under the same conditions as those used for deposition on the $SiO_2$ glass substrates. Details of the apparatus are given in our previous studies on SnS thin films [24,42]. The apparatus consists of a pressure-gradient sputtering (PGS) cathode (Kenix Co., Ltd., Japan) equipped with a 1-inch-diameter metallic target and a commercially available inductively coupled plasma (ICP)-type RF plasma cracking cell (Kenix Co., Ltd.). In the PGS cathode, Ar gas is introduced through the cathode into the vicinity of the target so that the pressure near the target is kept higher than that in the rest of the chamber. In the plasma cracking cell, elemental sulfur (99.99%, Kojundo Chemical Laboratory Co., Ltd., Japan) is heated to generate sulfur vapor, which is mixed with Ar gas (5 sccm) and activated by applying RF power (50 W) to the ICP coil. The distance from the outlet of the sulfur plasma source to the substrate was fixed at 85 mm, and the target-to-substrate distance was 135 mm. The metallic targets were Zn (99.99%, Mitsubishi Materials Corporation, Japan), Cu (99.99%, Mitsubishi Materials Corporation), W (99.9%, Plansee, Austria), and Mo (99.9%, Plansee). The base pressure of the deposition chamber was on the order of $10^{-6}$ Pa. During deposition, Ar gas was supplied to the sputtering cathode at 20 sccm, and the chamber pressure, measured near the exhaust port, was 0.5 Pa. The substrate temperature (TSub), the RF power applied to the metallic target (RFMetal), and the heating temperature of the elemental sulfur (TSulfur) for each sulfide are listed in Table 1. The deposition times were 30 min for the WS2 film, 1 h for the MoS2 film, 5 h for the ZnS and Cu1.8S film; the sulfide films used to evaluate the deposition rates in Table 2 were deposited for 5 h. To compare deposition rates, the corresponding metallic films were also deposited without heating the sulfur source or applying RF power to the ICP coil. During $MoS_2$ deposition with sulfur plasma, the rotary pump exhaust contained 0.3–0.4 ppm $H_2S$, while the $SO_2$ concentration was below the detection limit (0.02 ppm); even in the undiluted pump exhaust, the $H_2S$ concentration was below the occupational exposure limit set by the American Conference of Governmental Industrial Hygienists (ACGIH) (TLV-TWA for $H_2S$, 1 ppm [49]) (see Section S5 in the Supporting Information).

In the plasma cracking cell, sulfur vapor mixed with Ar gas is excited and dissociated by the ICP. The energy of the metastable excited states of Ar (11.55 eV [36]) exceeds the first ionization energy of S (10.36 eV [36]), and ionized sulfur species have been observed in similar $H_2S$/Ar discharges [50]. Some of the sulfur is therefore expected to be ionized in this apparatus as well, although the degrees of dissociation and ionization remain to be quantified. In this paper, the sulfur species supplied in this way are referred to as sulfur plasma.

**Characterization:** The crystal phases of the films were identified by X-ray diffraction (XRD; SmartLab, Rigaku Corporation, Japan) with Cu Kα radiation in the out-of-plane and in-plane geometries. The film compositions were determined by field-emission electron probe microanalysis (FE-EPMA; JXA-8530, JEOL Ltd., Japan). The surface and cross-sectional morphologies of the films were observed by field-emission scanning electron microscopy (FE-SEM; JSM-7800F, JEOL Ltd.). The film thicknesses were evaluated by cross-sectional FE-SEM and with a stylus profilometer (Dektak XT, Bruker, USA). The optical transmission spectra of the ZnS films were measured with an ultraviolet-visible-near-infrared spectrophotometer (U-4100, Hitachi High-Tech Corporation, Japan).

To directly observe the layered structure of the $WS_2$ film, a cross-sectional specimen was prepared by focused ion beam (FIB) milling (Helios, FEI/Thermo Fisher Scientific, USA) and examined by high-resolution transmission electron microscopy (HRTEM). HRTEM images were acquired using an aberration-corrected transmission electron microscope (JEM-ARM200F, JEOL Ltd.) operated at 200 kV. The film region in the cross-sectional HRTEM images was analyzed by fast Fourier transform (FFT) using ImageJ [51], and interplanar spacings were calculated from the positions of the resulting diffraction spots.

## Author Contributions

Daiki Motai: Investigation, Methodology, Formal analysis, Data curation, Writing – original draft. Issei Suzuki: Conceptualization, Formal analysis, Visualization, Supervision, Project administration, Funding acquisition, Writing – original draft. Taichi Nogami: Investigation, Methodology. Takahisa Omata: Supervision, Writing – review & editing.

## Acknowledgements

The authors are grateful to Prof. Takeo Nakano (Seikei University) for valuable discussions on the mechanism by which sulfur plasma promotes low-temperature crystallization. This work was partly supported by JSPS KAKENHI Grants-in-Aid for Scientific Research (B) (Grant Nos. JP21H01613 and JP26K01204) and a Grant-in-Aid for Challenging Research (Exploratory) (Grant No. JP26K22577), the JST FOREST Program (Grant No. JPMJFR256C), the Tsukuba Innovation Arena (TIA) "Kakehashi" collaborative research program, the Cooperative Research Program of the "Network Joint Research Center for Materials and Devices (MEXT)", and a research grant from the Proterial Materials Science Foundation. Part of this study was supported by the Research equipment sharing system, Tohoku University (072・FE-EPMA, 099・FE-SEM), and the Support system for young researchers to use research equipment, Tohoku University. This work used research equipment shared through the MEXT Project for promoting public utilization of advanced research infrastructure (Program for supporting construction of core facilities) (Grant No. JPMXS0440600025).

Supplementary Material for:

# Sulfur Plasma Opens Low-Temperature Reactive Sputtering to Diverse Sulfides

Daiki Motai[1], Issei Suzuki[1,*], Taichi Nogami[1], Takahisa Omata[1]

## Section S1. Lattice constants of the $WS_2$ and $MoS_2$ thin films

Table S1 lists the lattice constants of the $WS_2$ and $MoS_2$ films deposited at 300 °C, together with the bulk values. In both films, the c axis determined from the out-of-plane diffraction angle is expanded relative to the bulk, whereas the a axis determined from the in-plane diffraction angles is contracted. The c-axis expansion is particularly large for $WS_2$, at 13%. Such c-axis expansion has been widely reported for sputtered TMD films: +7% [1] and +4% to +25% [2] for $MoS_2$ films, and up to +3% for $WS_2$ films [3]. A contraction of the a axis accompanying the c-axis expansion has also been reported for sputtered $MoS_2$ films [2]. The c-axis expansion has been interpreted as an increase in the interlayer distance caused by turbostratic stacking, in which adjacent layers are slightly rotated and translated relative to regular 2H stacking [1,3]. For the $WS_2$ film in this work, the interlayer spacing derived from the 002 reflection agrees with the values obtained from the HRTEM image (7.0 Å) and the FFT pattern (7.1 Å), confirming the c-axis expansion in real space.

**Table S1.** Lattice constants *a* and *c* of the $WS_2$ and $MoS_2$ thin films deposited at 300 °C, determined from the 110 and 200 reflections in the in-plane XRD patterns and from the 002 reflection in the out-of-plane XRD patterns (Figure 2), respectively. The values for bulk 2H-$WS_2$ [4] and 2H-$MoS_2$ [5] are also listed. The values in parentheses indicate the deviations from the bulk values.

| | *a* / Å | *c* / Å |
|---|---|---|
| $WS_2$ thin film | 3.02 (−4.4%) | 14.00 (+13.3%) |
| 2H-$WS_2$ (bulk)[a] | 3.16 | 12.36 |
| $MoS_2$ thin film | 3.13 (−1.3%) | 12.86 (+4.4%) |
| 2H-$MoS_2$ (bulk)[b] | 3.17 | 12.32 |

[a]Ref.[4], [b]Ref.[5]

**Section S2. Morphology of the ZnS thin films**

Figure S1 shows surface and cross-sectional SEM images of the ZnS films presented in Figure 3 of the main text. Both films (S/Zn = 0.71 and 0.91) have flat surfaces without a distinct granular structure. The film thicknesses determined from the cross-sectional images are 350 nm and 440 nm, respectively, and both films are deposited on the substrates as uniform, dense layers.

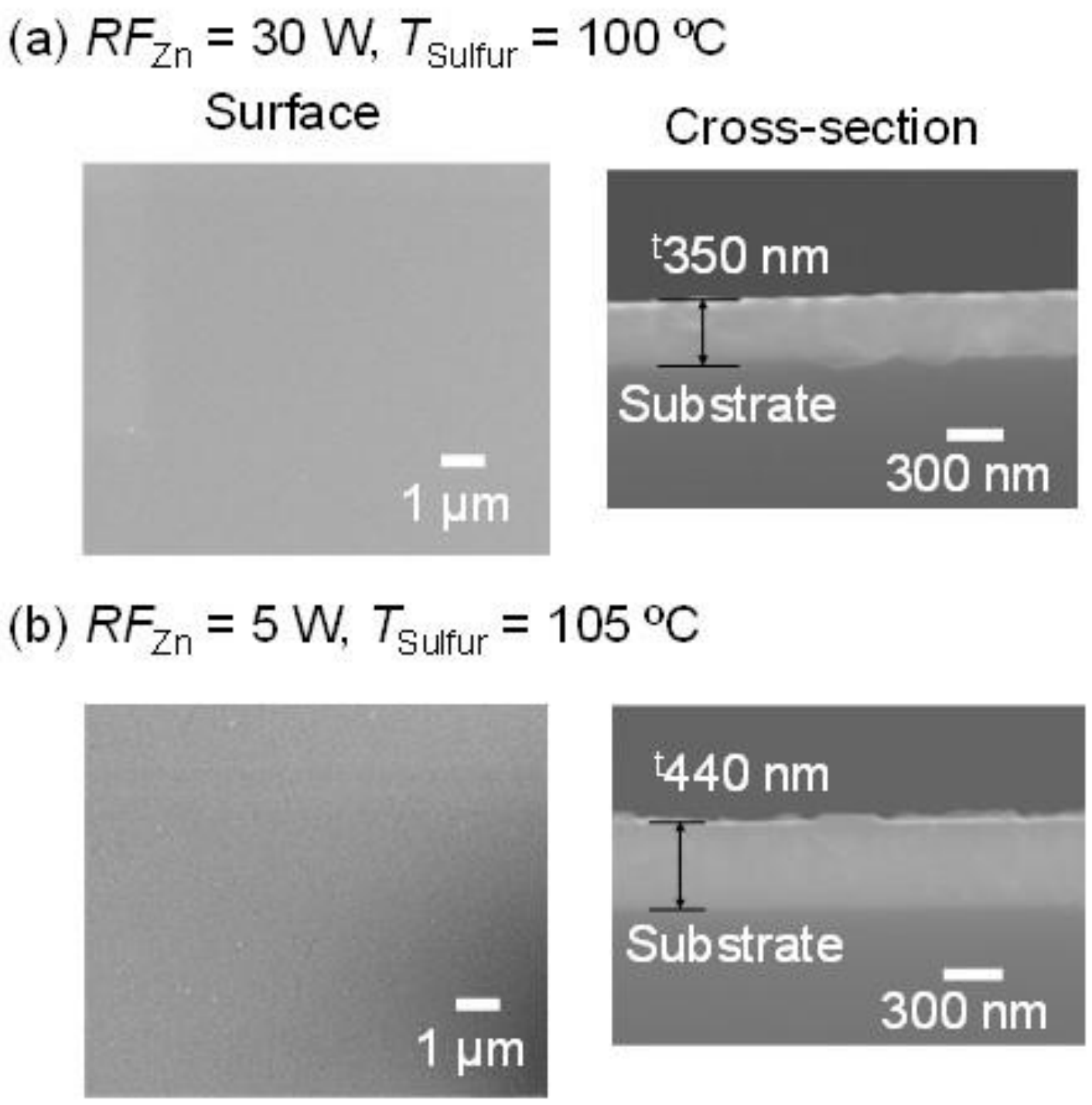

**Figure S1.** Surface (left) and cross-sectional (right) SEM images of the ZnS thin films deposited on unheated $SiO_2$ glass substrates (a) at $RF_{Zn}$ = 30 W and $T_{Sulfur}$ = 100 °C (S/Zn = 0.71), and (b) $RF_{Zn}$ = 5 W and $T_{Sulfur}$= 105 °C (S/Zn = 0.91).

## Section S3. ZnS thin film deposited without sulfur plasma

As shown in Figure 3 of the main text, wurtzite-type ZnS films were obtained under both conditions when a metallic Zn target was used with sulfur plasma. In contrast, the film sputtered from a ZnS target without sulfur plasma (TSub = 200 °C) shows reflections characteristic of both wurtzite-type and zincblende-type ZnS (Figure S2), indicating a mixture of the two phases.

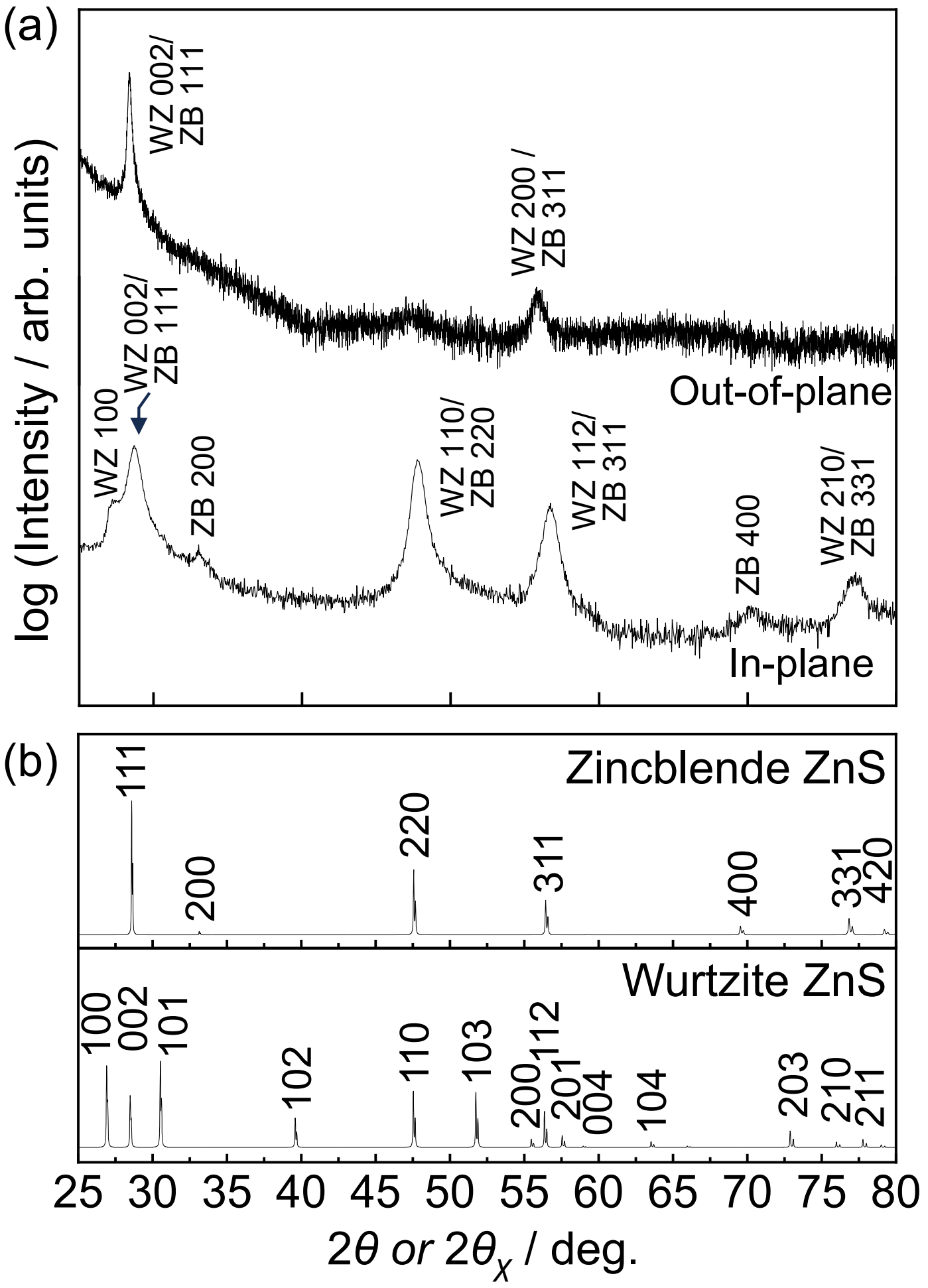

**Figure S2.** (a) Out-of-plane and in-plane XRD patterns of a ZnS thin film deposited from a ZnS target at TSub = 200 °C without sulfur plasma supply. (b) Powder patterns of wurtzite-type (ICSD#230703 [6]) and zincblende-type ZnS (ICSD#67453 [7]).

**Section S4. Fabrication of $Cu_{1.8}S$ thin film**

Figure S3(a) shows the out-of-plane and in-plane XRD patterns of the film deposited by reactive sputtering of the Cu metallic target with sulfur plasma (see Table 1 in the main text for the deposition conditions). All observed reflections match the powder pattern of digenite-type $Cu_{1.8}S$, confirming that the film consists of this phase. As shown in Figure S3(b,c), the film has a flat surface and a uniform thickness of approximately 300 nm.

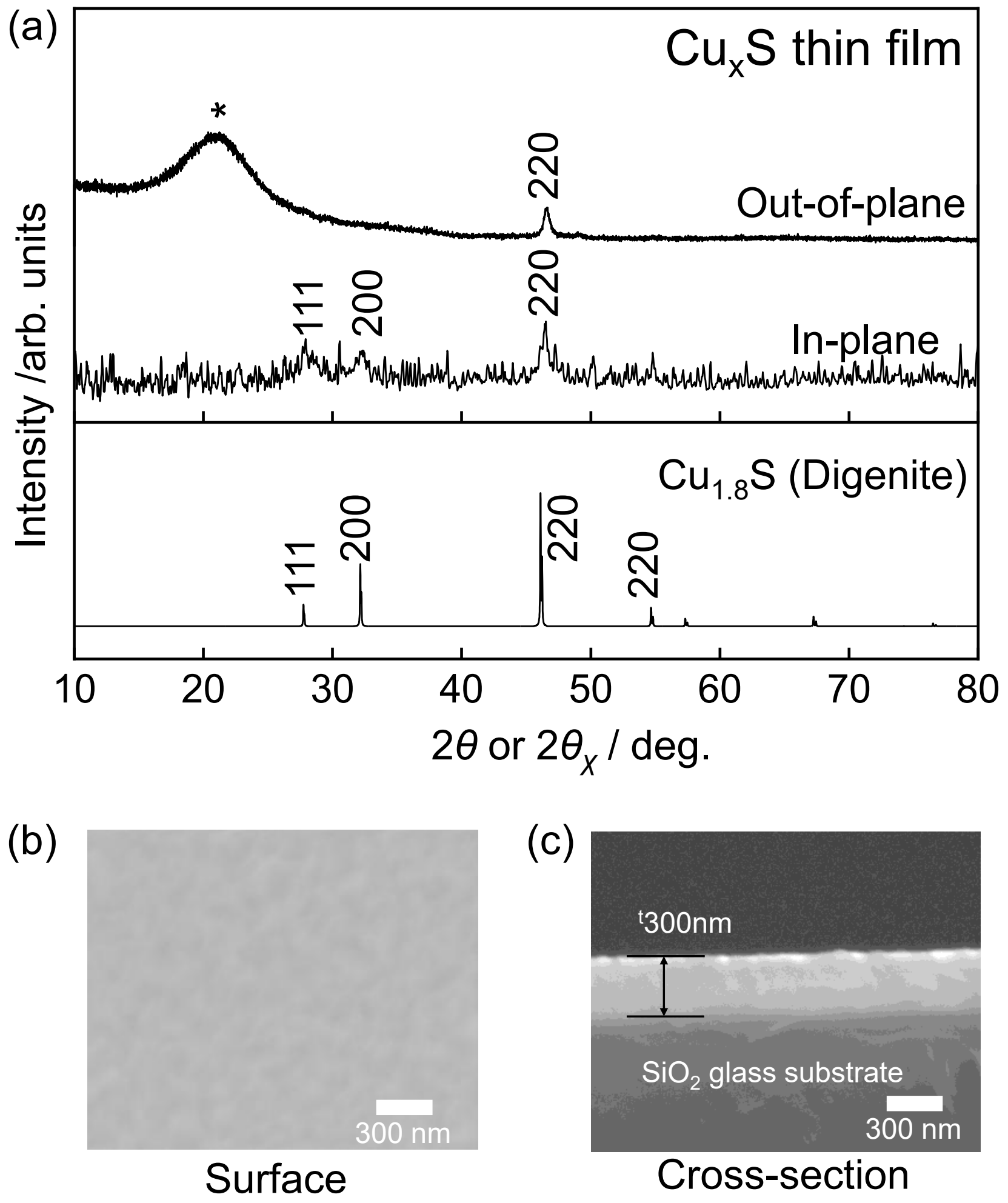

**Figure S3.** (a) Out-of-plane and in-plane XRD patterns of the $Cu_{1.8}S$ thin film, together with the powder pattern of digenite-type $Cu_{1.8}S$ (ICSD#41142 [8]). The broad peak marked with an asterisk originates from the $SiO_2$ glass substrate. The deposition time was 5 h. (b) Surface and (c) cross-sectional SEM images of the thin film.

**Section S5. Exhaust gas from the deposition process with sulfur plasma**

$H_2S$ and $SO_2$ in the exhaust gas were measured using gas detector tubes (GASTEC No. 4LT and No. 5LC, GASTEC Corporation, Japan; detection limits of 0.01 ppm and 0.02 ppm, respectively). During $MoS_2$ deposition (55 min after the start of deposition), 0.3–0.4 ppm $H_2S$ was detected in the rotary pump exhaust line, while $SO_2$ was below the detection limit. Even in the undiluted pump exhaust, this $H_2S$ concentration was below the occupational exposure limit (ACGIH TLV-TWA, 1 ppm [9]). Neither gas was detected above its respective detection limit when the same gas flows used during deposition were introduced without heating the sulfur source or applying RF power to the ICP coil, when no gas was introduced, or in the laboratory air.

**Section S6. Deposition rates of metal atoms in the metallic and sulfide films**

For Zn, the deposition rate of metal atoms in the sulfide film was about 1.7 times that in the metallic film, opposite to the trend for the other metals. The vapor pressure of metallic Zn is many orders of magnitude higher than those of Cu, Sn, Mo, and W [10]. Zn is known to have a low condensation coefficient at low deposition rates [11]. During deposition of the metallic film, some of the Zn arriving at the substrate is therefore likely to have re-evaporated rather than remaining in the film, lowering the deposition rate of the reference metallic film. Under sulfur plasma supply, in contrast, Zn is incorporated into the film as ZnS, which has a considerably lower vapor pressure, so re-evaporation is suppressed and the ratio exceeds 1.

**Table S2.** Deposition rates of metal atoms (i) in metallic films deposited without sulfur supply and (ii) in sulfide films deposited with sulfur plasma, together with the ratios (ii)/(i). The rates were estimated from film thicknesses measured by cross-sectional SEM, deposition times (5 h for the sulfide films and 1–5 h for the metallic films), and the bulk densities of the respective phases.

| Metal / Sulfide | $RF_{Metal}$ / W | (i) Deposition rate of metal atoms as metallic film/ atom·cm$^{-2}$·h$^{-1}$ | (ii) De$_{pos}$ition rate of metal atoms as sulfide / atom·cm$^{-2}$·h$^{-1}$ | (ii)/(i) |
|---|---|---|---|---|
| Cu / C$^{u}_{1.8}$S | 10 | 3.1×10$^{17}$ | 2.6×10$^{17}$ | ~0.8 |
| W / $WS_2$ | 5 | 6.1×10$^{16}$ | 4.0×10$^{16}$ | ~0.7 |
| Mo / M$^{oS}_{2}$ | 5 | 1.0×10$^{17}$ | 5.9×10$^{16}$ | ~0.6 |
| Sn / SnS[a] | 5 | 1.3×10$^{17}$ | 8.9×10$^{16}$ | ~0.7 |
| Zn / ZnS | 5 | 1.1×10$^{17}$ | 1.9×10$^{17}$ | ~1.7 |

[a]Ref. [12].

## Supplementary references